\documentclass{svproc}
\usepackage{url}

\usepackage{graphicx}
\usepackage[implicit=false]{hyperref}

\begin{document}
\mainmatter              
\title{An Improved Algorithm for Fast K-Word Proximity Search Based on Multi-Component Key Indexes}
\titlerunning{An Improved Algorithm for Fast K-Word Proximity Search}  
%
\author{Alexander B. Veretennikov\inst{1}}
\authorrunning{Alexander B. Veretennikov} 
%
\tocauthor{Alexander B. Veretennikov}
\institute{Ural Federal University, Yekaterinburg, Russia\\
Chair of Calculation Mathematics and Computer Science\\
\email{alexander@veretennikov.ru},\\ WWW home page:
\texttt{http://veretennikov.org}
}

\maketitle              

This is a pre-print of a contribution published in 
Arai K., Kapoor S., Bhatia R. (eds) Intelligent Systems and Applications. IntelliSys 2020. Advances in Intelligent Systems and Computing, vol 1251, published by 
Springer, Cham. The final authenticated version is available online at: 

\noindent
\href{https://doi.org/10.1007/978-3-030-55187-2\_37}{https://doi.org/10.1007/978-3-030-55187-2\_37}.

\begin{abstract}
A search query consists of several words. In a proximity full-text search, we want to find documents that contain these words near each other. This task requires much time when the query consists of high-frequently occurring words. If we cannot avoid this task by excluding high-frequently occurring words from consideration by declaring them as stop words, then we can optimize our solution by introducing additional indexes for faster execution. In a previous work, we discussed how to decrease the search time with multi-component key indexes. We had shown that additional indexes can be used to improve the average query execution time up to 130 times if queries consisted of high-frequently occurring words. In this paper, we present another search algorithm that overcomes some limitations of our previous algorithm and provides even more performance gain.

\end{abstract}

\section{Introduction}

\tolerance=6000

\begin{figure}[t]
  \centering
  \setlength{\abovecaptionskip}{1pt}
  \setlength{\belowcaptionskip}{1pt}
  \includegraphics[width=0.45\textwidth]{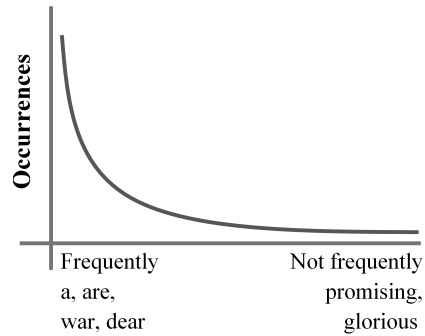}
  \caption{A typical word frequency distribution.}
  \label{VeretennikovA-imageZipf}
\end{figure}

Full-text search is a cornerstone of information retrieval. By a list of words, a user can obtain relevant documents that contain these words. Inverted files are used for this search \cite{Zobel:2006:IFT:1132956.1132959,Luk:2011:SSS:1988089.1988471,YangBlockLinked,BorodinGIN}. Words occur in documents with different frequencies, and an example of the frequency distribution of words is represented by Zipf's law \cite{Zipf:1929}, which is presented in Figure \ref{VeretennikovA-imageZipf}. On the horizontal axis, we plot words from high-frequently occurring to low-frequently occurring. On the vertical axis, we plot the numbers of occurrences of the corresponding words in a typical text collection.

The most frequently occurring words (see Figure \ref{VeretennikovA-imageZipf}, on the left side) occur significantly more often than ordinary words (see Figure \ref{VeretennikovA-imageZipf}, on the right side). This factor can affect the search performance in some cases. If the user needs only the document that contains the query words, then the search query time depends only on the number of documents in the collection. For each document and each word that occurs in the document, we need to store in the index exactly one record, which represents the fact of occurrence of the word somewhere in the document. 
 
For other kinds of full-text searches, we need to store a record for each occurrence of each word in each document \cite{Sadakane1,GallKWordEncrypted}, which considerably affects performance. In this case, the search time depends on the occurrence frequency in the texts of the queried words, and it is common to observe a search system that evaluates queries, which contain high-frequently occurring words in a significantly longer time than queries that consist only of ordinary words (see the left side and the right side of Figure \ref{VeretennikovA-imageZipf} respectively). See an example in \cite{Veretennikov:DAMDID:2018}. 

One way is to skip the most frequently occurring words. However, there are some concerns about this approach \cite{Williams:2004:FPQ:1028099.1028102}. A high-frequently occurring word may have a unique meaning in the context of the specific query. The authors \cite{Williams:2004:FPQ:1028099.1028102} stated literally that ``stopping or ignoring common words will have an unpredictable effect''. Examples are provided in \cite{Williams:2004:FPQ:1028099.1028102,Veretennikov:DAMDID:2018}. We can consider as an example the query ``Who are you who''. The word ``Who'' has a specific meaning in this query: The Who are an English rock band, and ``Who are You'' is one of their songs.

If the user needs the document that contains the query as a phrase, that is, the queried words must exist in the document in sequential order one after another, then additional phrase indexes can be used for performance improvement \cite{Williams:2004:FPQ:1028099.1028102}. However, the phrase indexes cannot be used for proximity full-text search, that is, when the user needs a document that contains queried words near each other. In the latter kind of searches, some other words are allowed in the text between queried words. We have proposed other methods to solve this task \cite{Veretennikov:DAMDID:2018,Veretennikov:IntelliSys:2018,Veretennikov:SouthUral:2018}.

In our methodology, we define several types of queries, depending on the kinds of words they contain. For each type of query, we can use specific types of additional indexes. The kinds of words are specified on the basis of word frequencies.

The importance of proximity full-text searches is determined by involving the proximity factor in modern information-retrieval methods \cite{Yan:2010:ETP:1871437.1871593,Buttcher:2006:TPS:1148170.1148285,Rasolofo:2003:TPS:1757788.1757808,Lu:2016:EEH:2970398.2970404}.

Early termination approaches \cite{Anh:2001:VRE:383952.383957,Garcia:2004:AI:979922.979924} can reduce the query processing time by sorting the posting lists in the index according to the relevance in decreasing order. In this case, irrelevant records, which are located at the end of each posting list, can be skipped. However, these methods cannot be used in an effective way when we need proximity full-text searches \cite{Veretennikov:SouthUral:2018}. When we are sort posting lists according to some factors, they are sorted independently of each other. However, in a specific query, we have several words linked together, and we cannot skip any part of any posting list because it is always a possibility that records for a document that contains queried words near each other occur at the end of a posting list, due to the document having low relevance according to the nonproximity factors. This problem was investigated in \cite{Yan:2010:ETP:1871437.1871593} but only for two-word queries, demonstrating to be a huge limitation.

In the following sections, we introduce several lemma types and several index types, the definition of which is based on the defined lemma types. Then, we provide an overview of previously developed search algorithms. Then, the new algorithm is described. Then, the results of the experiments is presented.

\section{Lemmatization and Lemma Types}

We use a morphological analyzer for lemmatization. For each word in the dictionary, the analyzer returns the list of lemmas, i.e., basic or canonical forms. 
Our dictionary now supports two languages.

Let us sort all lemmas in decreasing order of their occurrences in the texts. We call such a sorted list $FL$-list. 

Let the first $SWCount$ elements of $FL$-list be ``stop'' lemmas. 

Let the second $FUCount$ elements of $FL$-list be ``frequently used'' lemmas. 

Let all remaining lemmas be ``ordinary'' lemmas. 

$SWCount$ and $FUCount$ are parameters in which the representative example values are 700 and 2100. 

We use $FL$-numbers to establish an order in the set of all lemmas. For example, ``you'' $<$ ``who'' because ``you'' has $FL$-number 47, and ``who'' has $FL$-number 293.

The examples for each type of word are as follows:

stop lemmas: ``are'', ``war'', ``time'', ``be''.

frequently used lemmas: ``beautiful'', ``red'', ``hair''.

ordinary lemmas: ``glorious'', ``promising''.

Although we introduce the notion ``stop lemma'', we do not exclude such lemmas from the search. This division of lemmas is only performed to introduce different optimization methods for each kind of lemma.

\section{Index Type}

The expanded $(f, s, t)$ index or three-component key index \cite{Veretennikov:IntelliSys:2018,Veretennikov:SouthUral:2018} is the list of occurrences of the lemma $f$ for which lemmas $s$ and $t$ both occur in the text at distances less than or equal to $MaxDistance$ from $f$. 

We create the expanded $(f, s, t)$ index only when $f$, $s$, and $t$ are all stop lemmas and only for the case in which $f  \leq  s  \leq  t$. There, $MaxDistance$ is a parameter which may have a value of 5, 7, 9 or even more.

Each posting in the index includes the distance between $f$ and $s$ in the text and the distance between $f$ and $t$ in the text.

The expanded $(w, v)$ index or two-component key index \cite{Veretennikov:IntelliSys:2018,Veretennikov:SouthUral:2018} is the list of occurrences of the lemma $w$ for which lemma $v$ occurs in the text at a distance less than or equal to $MaxDistance$ from $w$. 

We create the expanded $(w, v)$ index only when $w$ is a frequently used lemma and $v$ is a frequently used or ordinary lemma. Each posting in the index includes the distance between $w$ and $v$ in the text. If both $w$ and $v$ are frequently used lemmas, then we create an index for them only if $w < v$.

An ordinary inverted index with NSW records \cite{Veretennikov:IntelliSys:2018,Veretennikov:SouthUral:2018} contains the posting lists for each frequently used and ordinary lemma. Each posting includes an NSW (near stop word) record. This record contains information about all stop lemmas that occur in the texts near the position of the specified posting. NSW records can also be skipped if it is required.

Different types of indexes can be used depending on the types of lemmas in the query.

For example, if the query contains several ordinary lemmas and a frequently used lemma, then $(w, v)$ indexes can be used instead an ordinary index for obtaining information about the occurrence in the texts of these lemmas.

If the query contains only stop lemmas, we use $(f, s, t)$ indexes, because two-component indexes do not provide enough performance \cite{Veretennikov:DAMDID:2018}. This case is the most complex from the performance point of view. We investigated other types of queries in our previous work \cite{Veretennikov:IntelliSys:2018}, and the task for them seems to be solved. In the current work, we investigate only queries that consist only of stop lemmas.

Let us consider an example. Let us have two documents $D0$ and $D1$. The words are numbered, and these numbers are zero based.

$D0$: Who are you is the album by The Who.

$D1$: Who has reality, who is real, who is true.

Stop lemmas: who, are, you, is, the, by, etc.

We have several three-component keys there, for example: (you, are, who), (have, who, who), (the, by, who), (the, you, are), (be, who, who), etc.

Let us note that ``be'' is the lemma of ``is'', and ``have'' is the lemma of ``has''.

For the key (be, who, who) we have, for example, the records in the index are as follows: 

\noindent
$(0, 3, -3, 5)$, $(1, 4, -4, -1)$, $(1, 4, -1, 2)$, $(1, 4, -4, 2)$, $(1, 7, -4, -1)$.

Let us consider $(0, 3, -3, 5)$, for example. There, 0 is the id of the document $D0$, 3 is the position of ``is'' in the document, $(-3)$ is the distance between the first ``who'' and ``is'', and 5 is the distance between the second ``who'' and ``is''.

For the key (you, are, who) we have, in the record $(0, 2, -1, -2)$, for example, 0 is the id of the document $D0$, 2 is the position of ``you'' in the document, $(-1)$ is the distance between ``are'' and ``you'' in the document, and $(-2)$ is the distance between ``you'' and ``who'' in the document.

\section{Types of Search Algorithms}

The expanded $(f, s, t)$ index contains $(ID, P, D1, D2)$ records, in which $ID$ is the identifier of the document, $P$ is the position of the word in the document, $D1$ is the distance between $s$ and $f$, and $D2$ is the distance between $t$ and $f$. 

If we have two records, \mbox{$A = (ID1, P1, X1, X2)$} and \mbox{$B = (ID2, P2, Y1, Y2)$}, then we define that $A < B$ if one of the following conditions is met: $ID1 < ID2$ or ($ID1 = ID2$ and $P1 < P2$). These records are stored in the index in increasing order. 

The iterator object can be used for reading all the records for the specific key. The iterator object has the $Next$ method with which we move to the next record. The iterator object also has the $Value$ property to access the current record. The iterator object also has the $Key$ property to access the key of the iterator. We can access specific components of the three-component key by $Key[0]$, $Key[1]$, and $Key[2]$.

We defined several search algorithms for multi-component key indexes.

In the \textit{Main-Cell} algorithm \cite{Veretennikov:SouthUral:2018}, we need to select the most frequently occurring lemma in the query. This lemma is called the main lemma of the query. Then, we form a list of multi-component keys. The main lemma is always the first component of each key. For other components, we are using the remaining lemmas of the query. 

We create an iterator object for each key. Then, in each iterator object, we use the $Next$ method to move to an equal position. After all iterator objects have equal position $(ID, P)$, we check that all lemmas are present nearby that position and calculate the size of the fragment of the text which contains the query. The drawback of this algorithm is that we need to duplicate the main lemma in several keys.

In the \textit{Intermediate-Lists} algorithm \cite{Veretennikov:DAMDID:2018}, we do not need to select a main lemma. We select a list of multi-component keys in such a way that each lemma of the query is used in some key. For each key, we have a list of records $(ID, P, D1, D2)$. From each record $(ID, P, D1, D2)$, we can create three records --- $(ID, P)$, $(ID, P + D1)$, $(ID, P+D2)$ --- that correspond to occurrences in the texts of $f$, $s$ and $t$ accordingly. 

The algorithm works as follows. We move in each iterator object to the same document. Then, for each iterator, we read all records for this document and produce three intermediate streams of records. Each intermediate stream contains a list of occurrences of a specific lemma in the document. Then, we combine the intermediate streams to produce results. The drawback of this algorithm is that we need to produce intermediate streams.

In the \textit{Optimized-Intermediate-Lists} algorithm \cite{Veretennikov:DAMDID:2019}, we are obtaining more performance gain by applying optimized key selection methods. But we still need to produce intermediate posting lists with this algorithm.

In this paper, we present a novel algorithm in which we combine the several posting lists for multi-component keys into a list of results without creating intermediate posting lists. We call this algorithm $Combiner$ algorithm.

\section{The Search Algorithm}

Let us have subquery $Q$, which is a list consisting of $n$ lemmas.

\begin{figure}[t]
  \centering
  \setlength{\abovecaptionskip}{1pt}
  \setlength{\belowcaptionskip}{1pt}
  \includegraphics[width=0.85\textwidth]{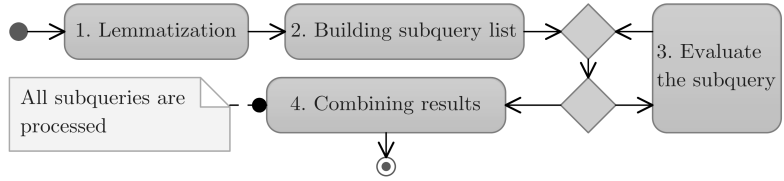}
  \caption{The search algorithm.}
  \label{VeretennikovA-imageSearch}
\end{figure}

The search algorithm consists of the following steps (see Figure \ref{VeretennikovA-imageSearch}).

\begin{itemize}
\item[1)]	Lemmatization
\item[2)]	Building the list of subqueries.
\item[3)]	Processing subqueries.
\item[4)]	Combining results.
\end{itemize}

Let us consider the following query: ``who are you who''. After lemmatization, we have [who] [are, be] [you] [who], because the word ``are'' has two lemmas in our dictionary, namely, ``are'' and ``be''.

For a query that consists of high-frequently occurring words, we need to create subqueries, that is, [who] [are] [you] [who] and [who] [be] [you] [who]. We need to have a subquery in the form of a list of lemmas. Then, we evaluate each subquery and combine the results.

\begin{figure}[t]
  \centering
  \setlength{\abovecaptionskip}{1pt}
  \setlength{\belowcaptionskip}{1pt}
  \includegraphics[width=0.95\textwidth]{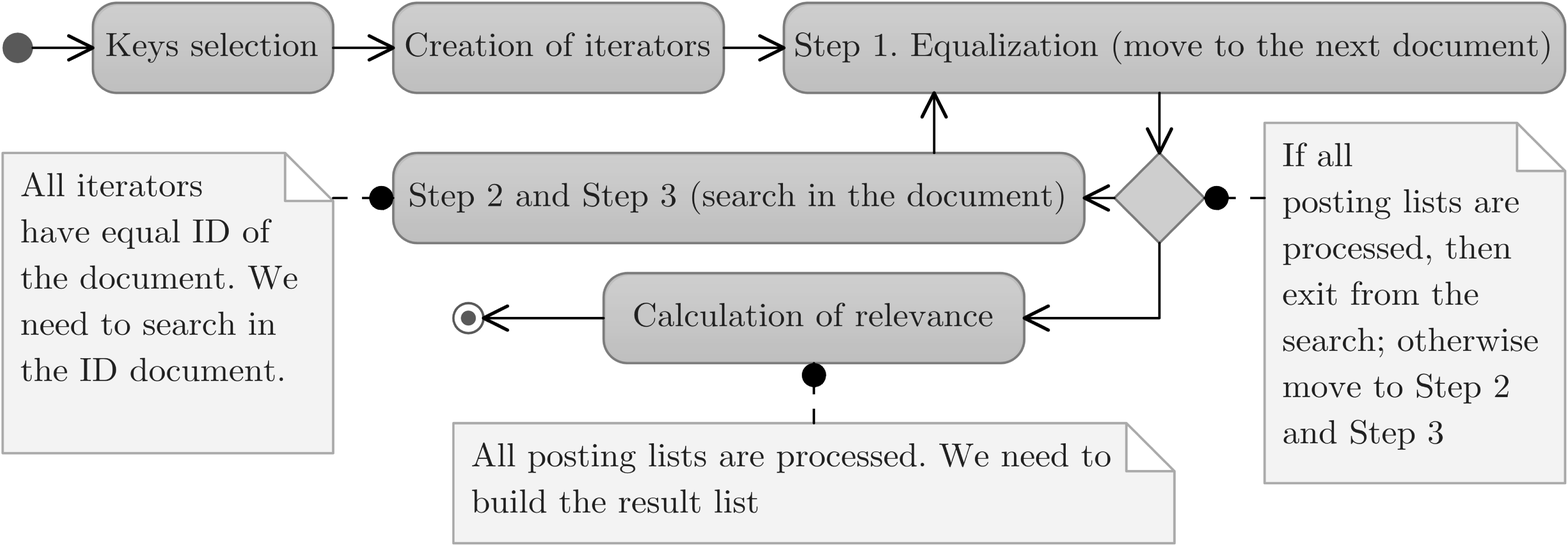}
  \caption{The processing of the evaluation of the subquery.}
  \label{VeretennikovA-imageSearchSub}
\end{figure}

The processing of the evaluation of the subquery contains the following stages (see Figure \ref{VeretennikovA-imageSearchSub}).

\begin{itemize}
\item[1)]	Selection of the keys.
\item[2)]	Building an iterator for each key.
\item[3)]	Search.
\item[4)]	Calculation of the relevance.
\end{itemize}

\section{Key Selection}

We have a subquery that is a list of lemmas. The elements of the list are numbered starting with zero. For each key, we need to select three components, and we need to select those using lemmas that occur at different indexes in the subquery. We also want to exclude duplicates from consideration, if the subquery has some lemmas that appear several times. When we select a lemma as a component of a key, we ``mark'' it as ``used''. Let $Used$ be a set of lemmas that is initially empty. When we mark a lemma, we include it into $Used$. We will use occurrence frequency in the texts of the lemmas as a factor for selection. 

For the first component of the first key, we select the most frequently occurring unused lemma in the query. 

For the second component of the key, we need to select an unused lemma in the query in which the index in the query is different from the index in the query of the first component of the key. If we have several acceptable lemmas, we select one among them that is the least frequently occurring in the texts. If we do not have any acceptable lemma, then we select a lemma using the aforementioned conditions, except we ignore the ``used'' mark, and we mark this component with * to designate it as duplicate. 

For the third component of the key, we need to select an unused lemma in the query in which the index in the query is different from the indexes in the query of the first and the second components of the key. If we have several acceptable lemmas, we select one among them that is the least frequently occurring in the texts. If we do not have any acceptable lemma, then we select a lemma using the aforementioned conditions except we ignore the ``used'' mark, and we mark this component with * to designate it as duplicate.

Then, we mark all selected lemmas as ``used''.

If we have any unused lemmas, we repeat the process and form another key; otherwise, all keys are selected.

Let us consider an example. Let us say that we have a query ``Who are you and why did you say what you did''. This query can be found in Cecil Forester Scott's novel ``Lord Hornblower''. Let us consider its subquery [who] [are] [you] [and] [why] [do] [you] [say] [what] [you] [do]. With $FL$-numbers, the query will have the following appearance: [who: 293] [are: 268] [you: 47] [and: 28] [why: 528] [do: 154] [you: 47] [say: 165] [what: 132] [you: 47] [do: 154].

We select ``and: 28'' as the first component of the first key, because it is the most frequently occurring lemma; that is, it has the least $FL$-number (28) among the lemmas of the subquery. Then, we select ``why: 528'' as the second component of the key and ``who: 293'' as the third component of the key. We mark ``and'', ``why'' and ``who'' as used.

We have other unused lemmas and can select another key. We select ``you: 47'' as the first component of the second key, and ``are: 268'' and ``say: 165'' as the second and the third components of the second key. We mark all selected lemmas as ``used''.
Then, we select ``what: 132'' as the first component of the third key. We select ``do: 154'' as the second component of the key. There are no ``unused'' lemmas remaining. Therefore, we ignore the ``used'' mark and select ``why*: 528'' as the third component of the third key. This component we mark with * because it is a duplicate.

\section{Search for a Subquery}

Queries are usually evaluated by Term-At-A-Time (TAAT), Document-At-A-Time (DAAT) or Score-At-A-Time (SAAT) approaches \cite{Daoud:2016:FTP:2978438.2978526,Jiang:2015:TEI:2839534.2840112}. DAAT approaches have advantages over SAAT and TAAT approaches \cite{Jiang:2015:TEI:2839534.2840112}.

We use a Document-At-A-Time (DAAT) kind of algorithm. An iterator allows reading the posting list for a key from the start to the end. The posting list is sorted in increasing order.

\begin{figure}[t]
  \centering
  \setlength{\abovecaptionskip}{1pt}
  \setlength{\belowcaptionskip}{1pt}
  \includegraphics[width=\textwidth]{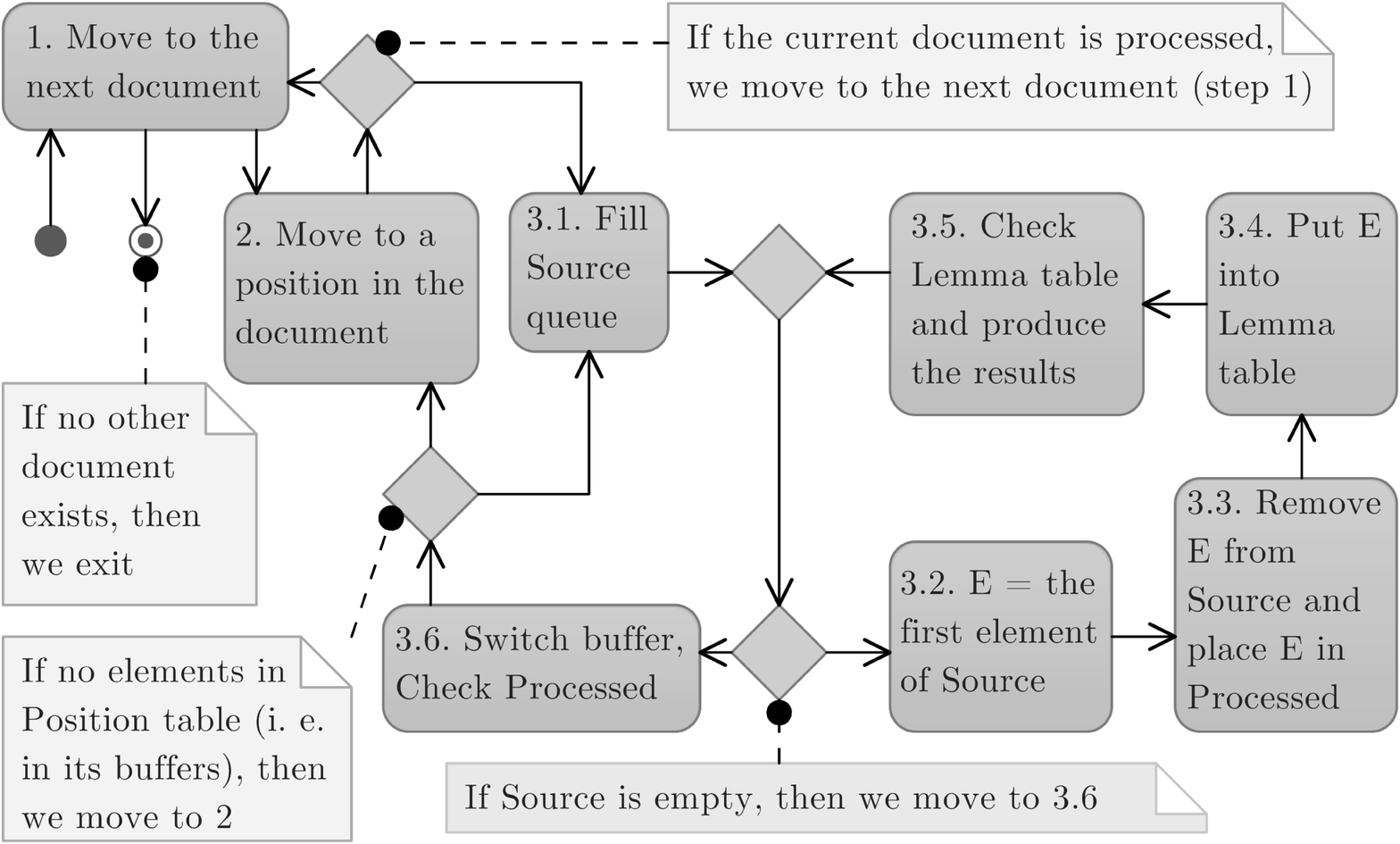}
  \caption{Search procedure diagram.}
  \label{VeretennikovA-imageAlgorithm}
\end{figure}

The search procedure is a three-level process (see Figure \ref{VeretennikovA-imageAlgorithm}):

Step 1. We move to a document. All iterators are positioned on the specific document.

Step 2. We move to a position in the document at which the queried lemmas are near each other.

Step 3. We put information about the position of the lemmas in special tables. We use the tables to check that all queried lemmas are present and to calculate the exact position of the result in the text.

If we do not have an acceptable position in the document in Step 2, then we move to the next document (move to the start of Step 1); otherwise, we repeat it.

If we do not have another document in Step 1, then we exit from the search; otherwise, we repeat it.
 
\section{Step 1}

If we have read and processed all postings, then we exit from the search; otherwise, we perform the following in a loop:

\begin{itemize}
\item[1)]	Let $S$ be the iterator with the minimum document identifier, which is defined by its $Value.ID$. 
\item[2)]	Let $E$ be the iterator with the maximum document identifier.
\item[3)]	If $S.ID = E.ID$, then we exit from the loop and move to Step 2; otherwise, we perform $S.Next()$ and move to the start of the loop again. 
\end{itemize}

The cost of one iteration of this loop is $O(\texttt{log } n)$. Both $S$ and $E$ are local variables for this procedure.

\section{Step 2}

We perform the search in the document $ID$. We perform in a loop the following.

\begin{itemize}
\item[1)]	If we have read all postings for the document $ID$ in some iterator, we break the loop; otherwise, we perform the following sub steps 2-4.
\item[2)]	Let $S$ be the iterator with the minimum position value $Value.P$. 
\item[3)]	Let $E$ be the iterator with the maximum position value. Let $Delta = E.Value.P - S.Value.P$. 
\item[4)]	If $Delta < MaxDistance  \times  2$, then break the loop and move to Step 3; otherwise, we perform $S.Next()$ and move to the start of the loop again. 
\end{itemize}

The cost of one iteration of this loop is $O(\texttt{log } n)$. We can use binary heaps \cite{Williams:HeapSort} to implement this approach.

Step 2 ends when one of the following cases occurs:

\begin{itemize}
\item[1)]	After execution of $S.Next$, we move to another document in $S$, or we do not have any other postings in $S$. In this case, we break the loop and move to Step 1.
\item[2)]	We have $Delta < MaxDistance  \times  2$. In this case, we have a place in the document that can potentially contain all queried lemmas near each other. Then, we break the loop and move to Step 3.
\end{itemize}

\section{Step 3}

At the start of this step, we have established that in the current position in the document, we have all keys, meaning that all queried lemmas are near each other. We have two tables, namely, $Lemma$ table and $Position$ table. We will use these tables to obtain the fragments of the text that contain the queried lemmas. Afterward, we move to Step 2 again.

\subsection{The $Lemma$ Table}

The $Lemma$ table is used for the following.

\begin{itemize}
\item[1)]	To check whether all queried lemmas exist in the text or not.
\item[2)]	To determine the start and the end of the fragment of the text that contains the queried lemmas. The fragment of the text must have the minimum length among the acceptable fragments.
\end{itemize}

The $Lemma$ table contains an array that consists of $SWCount$ entries. 

Each entry corresponds to one stop lemma by its $FL$-number. 

Each entry has two fields, that is, $Max$ and $Count$. 

$Max$ is the count of the occurrences of the corresponding lemma in the query. We initialize this field at the start of the search. $Count$ is the count of occurrences of the corresponding lemma in the current fragment of the text, which is initially zero.

The $Lemma$ table itself has $Max$ and $Count$ fields. 

The $Max$ field of the $Lemma$ table equals to the length of the subquery. We initialize this field at the start of the search.

The idea of the use of the $Lemma$ table is the following. We need two queues.

Let us have a queue of records $(P, Lem)$, where the lemma $Lem$ is some lemma from the subquery and $P$ is the position of the lemma $Lem$ in the document. This queue we call $Source$. Let the queue be sorted in increasing order of $P$. 

Let us also have a second queue. The second queue we call $Processed$. This queue will also be sorted in increasing order of $P$.

We will process all elements of the $Source$ from the start to the end. 

We perform the following in a loop, until we have any element in $Source$:

\begin{itemize}
\item[1)]	Let $E$ be the first, that is, the minimum, element from the $Source$ queue (see 3.2 in Figure \ref{VeretennikovA-imageAlgorithm}).
\item[2)]	We remove $E$ from the $Source$ queue (see 3.3 in Figure \ref{VeretennikovA-imageAlgorithm}).
\item[3)]	We place $E$ into the $Processed$ queue into the end of the queue (see 3.3 in Figure \ref{VeretennikovA-imageAlgorithm}).
\item[4)]	We will add information about $E$ into the $Lemma$ table (see 3.4 in Figure \ref{VeretennikovA-imageAlgorithm}). This information includes the following (4.a --- 4.c). 
\begin{itemize}
\item[a.]	 We obtain the entry $Entry$ by the value of $E.Lem$. 
\item[b.]	 If $Entry.Max > Entry.Count$, then we increment the $Lemma.Count$ field of the $Lemma$ table.
\item[c.]	 We increment $Entry.Count$. 
\end{itemize}
\item[5)]	We check the $Lemma$ table (3.5 in Figure \ref{VeretennikovA-imageAlgorithm}).
\end{itemize}

\subsection{Checking the $Lemma$ Table (Step 3.5 in Figure \ref{VeretennikovA-imageAlgorithm})}

If $Lemma.Count \neq Lemma.Max$, then we do nothing. 

Otherwise, $Lemma.Count = Lemma.Max$. Then, we have in the text all required lemmas.

Then, we need to obtain the minimum fragment of the text, which contains the queried lemmas. 

For this, we will use $Processed$ queue.

We perform in a loop the following:

\begin{itemize}
\item[1)]	Let $S$ be the first element of $Processed$ queue.
\item[2)]	We obtain the entry $Entry$ by the value of $S.Lem$.
\item[3)]	If $Entry.Count > Entry.Max$, then we can decrease the length of the fragment of the text. In this case, we decrease $Entry.Count$, remove $S$ from $Processed$, and go to 1; otherwise, we break the loop.
\end{itemize}

When we exit from the loop, $S$ defines the start of the fragment of the text, and $E$ defines the end of the fragment of the text.

The next question is where we obtain the $Source$ queue and how do we perform the sorting of $Source$ with $O(1)$ computational complexity.

\subsection{$Position$ Table}

The $Position$ table has the method $Set(P, Lem)$ where $P$ is the position of the lemma $Lem$ in the document. It has the property $Start$ which specifies the start of the current fragment of the text that is interesting for us. It has the method $Shift(P)$ which can be used to set the value of $Start$.

At the start of Step 3, we execute the method 
$$Shift(P - min(P, MaxDistance))$$ where $P$ is the minimum current position value $Value.P$ among all iterators.

In the internal implementation of the $Position$ table, we use three buffers each with a length $WindowSize$. Each buffer is an array which contains $WindowSize$ entries. The following condition must be met:
\begin{displaymath}
MaxDistance  \times  2  \leq  WindowSize  \leq  64.
\end{displaymath}

Each buffer also has a corresponding 64-bit $Mask$ field. Each entry of the buffer has a corresponding bit in the $Mask$ field of the buffer.

Each entry of the buffer has three fields: $Lem$, $P$, and $Next$.

When we execute $Set(P, Lem)$, we calculate the relative position in the $Position$ table, that is, 
\begin{displaymath}
R = P - Start. 
\end{displaymath}

Then, we define the buffer by performing $R / WindowSize$. 

Then, we define a relative position in the buffer 
\begin{displaymath}
RelativeP = R \: \% \: WindowSize 
\end{displaymath}
(let \% be the modulus operator).

The variable $RelativeP$ defines the target entry $T$ in the buffer. We set $T.Lem = Lem$ and $T.P = P$ for the target entry. We also set the bit with number $RelativeP$ in the $Mask$ field of the buffer to 1.

From the buffer, we can produce a queue, which is a sorted linked list.

We can use the Bit Scan Forward operation, which is one processor command, on the $Mask$ field to determine the index of the first entry of the queue. We can reset the bit of the entry to zero and perform the Bit Scan Forward operation again to move to the second entry and so on. To build the queue, we use $Next$ fields of entries. The queue will be initially sorted by this creation process.

The problem here is that one buffer has a limited length. To solve this problem, we use three buffers.

Let $WindowFlushBorder$ be $WindowSize  \times  1.5$, that is, the center of the second buffer.

\subsection{The Search Procedure with Three Buffers}

For each iterator $IT$, we perform the following. 

We execute the methods:
  \begin{displaymath}
 Set(IT.Value.P, IT.Key[0]), 
 \end{displaymath}
  \begin{displaymath}
 Set(IT.Value.P + IT.Value.D1, IT.Key[1]), 
 \end{displaymath}
  \begin{displaymath}
 Set(IT.Value.P+ IT.Value.D2, IT.Key[2]) 
 \end{displaymath}
  and perform $IT.Next()$. All these actions we perform until 
 \begin{displaymath}
 IT.Value.P < Start + WindowFlushBorder. 
 \end{displaymath}
 See 3.1 in Figure \ref{VeretennikovA-imageAlgorithm}. 

We also take in consideration (*) marks of each component of $IT.Key$. We perform $Set$ only for these components that do not have the (*) mark. That is, if the third component has the (*) mark, we do not perform $Set(IT.Value.P+ IT.Value.D2, IT.Key[2])$.

Each call of $Set$ defines an occurrence $P$ of a lemma in the document. That means for each three-component key, we may produce up to three values, and each of them defines an occurrence of a lemma in the document. We are sure that all values with condition $P < Start + WindowSize$ are already produced. 

This is because we have $WindowSize  \geq  MaxDistance  \times  2$, and for any iterator $IT$, all records with condition 
\begin{displaymath}
IT.Value.P < Start + WindowFlushBorder
\end{displaymath}
are already processed. 

For any iterator $IT$, any next record 
\begin{displaymath}
IT.Value.P  \geq  Start + WindowFlushBorder = 
Start + WindowSize  \times  1.5  \geq 
\end{displaymath}
\begin{displaymath}
Start + WindowSize + MaxDistance.
\end{displaymath}
Therefore, 
\begin{displaymath}
IT.Value.P + IT.Value.D1  \geq  Start + WindowSize, 
\end{displaymath}
because we have:
\begin{displaymath}
-MaxDistance  \leq  IT.Value.D1  \leq  MaxDistance.
\end{displaymath}

For $IT.Value.D2$, we have the same.

After we process all the iterators, we put all updated entries from the first buffer in the $Source$ queue (we use Bit Scan Forward for this). 

This completes step 3.1 in Figure \ref{VeretennikovA-imageAlgorithm}.

Then, we use the $Lemma$ table to produce a list of search results (see 3.2-3.5 in Figure \ref{VeretennikovA-imageAlgorithm}). Each search result is a fragment of the document which contains the queried lemmas. In this production process, all elements of the $Source$ queue will be processed. However, after the processing, some entries may remain in the $Processed$ queue.

\subsection{Buffer Switch (See 3.6 in Figure \ref{VeretennikovA-imageAlgorithm})}

Let us note that if any item exists in the $Processed$ queue, then this item can only belong to the first buffer.

After the $Source$ queue is processed, we can go to the start of Step 3 again. We renumber the buffers in a cyclic way. The first buffer we make the third buffer; the second buffer will be the first buffer; and the third buffer will be the second buffer.

We require three buffers because of these entries in the $Processed$ queue. We cannot reuse these entries again, until they remain in the $Processed$ queue. However, we have no problems here. For any iterator $IT$, after the buffer switch, we will read all records with the condition: $IT.Value.P < Start_{new} + WindowFlushBorder$. That means, these records will affect the entries only in the first and second buffers, which means the new third buffer, that is, the former first buffer, will not be affected.

We also remove each entry $Entry$ from the $Processed$ queue with the following condition: 
\begin{displaymath}
(Start + WindowSize - Entry.P) > MaxDistance  \times  2.
\end{displaymath}

The following entries that can be added into the $Processed$ queue in the next iteration of Step 3 can lie only in the new first buffer, that is, the former second buffer. These following entries will be far from such entries which we remove, so we can safely free them. 

In fact, we need to free them, because it can be that no records will be added into $Processed$ in the next iteration, and this cleaning of the $Processed$ queue ensures that no item will exist in the $Processed$ queue that belongs to the first or the second buffer at the start of Step 3.

Finally, we set $Start = Start + WindowSize$.

\subsection{$Lemma$ Table Renumbering}

To reduce the size of the $Lemma$ table, we can assign a local number, with respect to the subquery, for each lemma. In this case, the size of the $Lemma$ table will be equal to the count of unique lemmas in the subquery. 

\section{Experiment 1}

In our experiments, we use the collection of texts from \cite{Veretennikov:IntelliSys:2018,Veretennikov:SouthUral:2018,Veretennikov:DAMDID:2018} which consists of approximately 195 000 documents of plain text, fiction and magazine articles with a total size of 71.5 GB. 
The average document text size is approximately 384.5 KB.
In our previous experiments, we used $MaxDistance = 5$, $SWCount = 700$, and $FUCount = 2100$. 

We need to use the same parameters to perform a comparison between the algorithms. The search experiments were conducted using the experimental methodology from \cite{Veretennikov:SouthUral:2018}. We assume that in typical texts, the words are distributed similarly, as Zipf stated in \cite{Zipf:1929}. Therefore, the results obtained with our text collection will be relevant to other collections.

We used the following computational resources:

CPU: Intel(R) Core(TM) i7 CPU 920 @ 2.67 GHz.

HDD: 7200 RPM. RAM: 24 GB.
OS: Microsoft Windows 2008 R2 Enterprise.

We created the following indexes.

$Idx1$: the ordinary inverted index without any improvements, such as NSW records \cite{Veretennikov:SouthUral:2018}. The total size is 95 GB.

$Idx2$: our indexes, including the ordinary inverted index with the NSW records and the $(w, v)$ and $(f, s, t)$ indexes, where $MaxDistance = 5$. The total size is 746 GB.

Please note that the total size of each type of index includes the size of the repository (indexed texts in compressed form), which is 47.2 GB.

The size of $Idx2$ is considerable, but we assume that we can allocate some disk space if the performance is required.

We performed 975 queries, and all queries consisted only of stop lemmas. The query set was also selected as in \cite{Veretennikov:IntelliSys:2018,Veretennikov:SouthUral:2018,Veretennikov:DAMDID:2018}. All searches were performed in a single program thread. The query length was from 3 to 5 words. 

Jansen et al. \cite{Jansen:2000:RLR:342495.342498} have shown by analysis of the query logs of a search system that for a typical search system, the users use short queries with a length less than or equal to 5. Queries with a length of 6, for example, represented approximately 1\% of all queries in \cite{Jansen:2000:RLR:342495.342498}.

We performed the following experiments.

$SE1$: all queries are evaluated using the standard inverted index $Idx1$.

$SE2.1$: all queries are evaluated using $Idx2$ and the Main-Cell algorithm from \cite{Veretennikov:SouthUral:2018}.

$SE2.2$: all queries are evaluated using $Idx2$ and the Intermediate-Lists algorithm from \cite{Veretennikov:DAMDID:2018}.

$SE2.3$: all queries are evaluated using $Idx2$ and the Optimized-Intermediate-Lists algorithm from \cite{Veretennikov:DAMDID:2019}.

$SE2.4$: all queries are evaluated using $Idx2$ and the new algorithm presented in this paper.

Average query times:
$SE1$: 31.27 sec., $SE2.1$: 0.33 sec.,  $SE2.2$: 0.29 sec.,  $SE2.3$: 0.24 sec., and $SE2.4$: 0.22 sec.

Average data read sizes per query:
$SE1$: 745 MB, $SE2.1$: 8.45 MB, $SE2.2$: 6.82 MB, $SE2.3$: 6.16 MB, and $SE2.4$: 6.2 MB.

Average numbers of postings per query:
$SE1$: 193 million, $SE2.1$: 765 thousand, $SE2.2$: 559 thousand, $SE2.3$: 419 thousand, and $SE2.4$: 423 thousand.

\begin{figure}[t]
  \centering
  \setlength{\abovecaptionskip}{1pt}
  \setlength{\belowcaptionskip}{1pt}
  \includegraphics[width=0.45\textwidth]{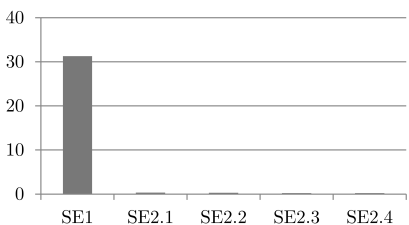}
  \caption{Average query execution times for $SE1$, $SE2.1$, $SE2.2$, $SE2.3$ and $SE2.4$ (seconds).}
  \label{VeretennikovA-ImageE1}
\end{figure}

We improved the query processing time by a factor of 94.7 with the $SE2.1$ algorithm, by a factor of 107.8 with the $SE2.2$ algorithm, by a factor of 130 with the $SE2.3$ algorithm, and by a factor of 142.13 with the $SE2.4$ algorithm, in comparison with ordinary inverted files $SE1$ (see Figure \ref{VeretennikovA-ImageE1}).
 
Let us consider Figure \ref{VeretennikovA-ImageE1}. The left-hand bar shows the average query execution time with the standard inverted indexes. The subsequent bars show the average query execution times with our indexes with the $SE2.1$, $SE2.2$, $SE2.3$ and $SE2.4$ algorithms. Our bars are much smaller than the left-hand bar because our searches are very fast. 
 
\begin{figure}[t]
  \centering
  \setlength{\abovecaptionskip}{1pt}
  \setlength{\belowcaptionskip}{1pt}
  \includegraphics[width=0.45\textwidth]{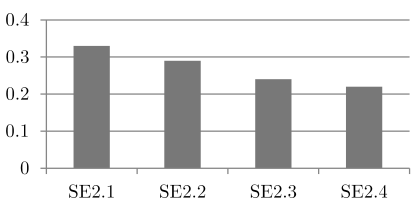}
  \caption{Average query execution times for $SE2.1$, $SE2.2$, $SE2.3$ and $SE2.4$ (seconds).}
  \label{VeretennikovA-ImageE2}
\end{figure}

We improved the query processing time by a factor of 1.09 with the $SE2.4$ algorithm in comparison with the $SE2.3$ algorithm, by a factor of 1.1 in comparison with the $SE2.2$ algorithm, and by a factor of 1.5 in comparison with the $SE2.1$ algorithm (see Figure \ref{VeretennikovA-ImageE2}).
 
\begin{figure}[t]
  \centering
  \setlength{\abovecaptionskip}{1pt}
  \setlength{\belowcaptionskip}{1pt}
  \includegraphics[width=1\textwidth]{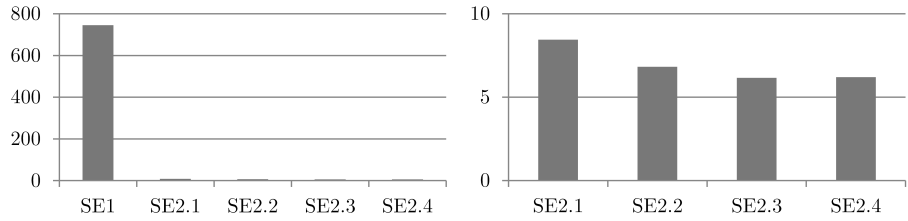}
  \caption{Experiment 1. Average data read sizes per query for $SE1$, $SE2.1$, $SE2.2$, $SE2.3$ and $SE2.4$ (MB).}
  \label{VeretennikovA-ImageE3-E4}
\end{figure}

We improved the average data read size per query by a factor of 88 with $SE2.1$, by a factor of 109.2 with $SE2.2$ and by a factor of 120 with $SE2.3$ and $SE2.4$, in comparison with ordinary inverted files $SE1$ (see Figure \ref{VeretennikovA-ImageE3-E4}).

We improved the average data read size per query by a factor of 1.1 with the $SE2.4$ algorithm in comparison with the $SE2.2$ algorithm and by a factor of 1.36 in comparison with the $SE2.1$ algorithm (see Figure \ref{VeretennikovA-ImageE3-E4}).

\section{Experiment 2}

For the second experiment GOV2 text collection and the following queries are used:
title queries from TREC Robust Task 2004 (with 250 queries in total), TREC Terabyte Task from 2004 to 2006 (with 150 queries in total) and TREC Web Task from 2009 to 2014 (with 300 queries in total), with 700 queries in total.
GOV2 text collection consists of approximately 25 million documents with a total size of approximately 426 GB, that is approximately 167 GB of plain text (after HTML tags removal). The average document text size is approximately 7 KB. We used $MaxDistance = 5$, $SWCount = 500$, and $FUCount = 1050$ and only English dictionary.
The value of $SWCount$ is very near to the 421 from \cite{Fox:1989:SLG:378881.378888}.

We created the following indexes.

$Idx1$: the ordinary inverted index without any improvements, such as NSW records \cite{Veretennikov:SouthUral:2018}. The total size is 143 GB (included the total size of indexed texts in compressed form, that is 57.3 GB).

$Idx2$: our indexes, including the ordinary inverted index with the NSW records and the $(w, v)$ and $(f, s, t)$ indexes, where $MaxDistance = 5$. The total size is 1.29 TB.

The query set can be divided into the following groups depending on lemmas in a concrete query:

\begin{enumerate}
\item[Q1)] Only stop lemmas: 12 queries.
\item[Q2)] Stop and frequently used and/or ordinary lemmas (i. e. the query has one or several stop lemmas and some other lemmas 
that may be frequently used or ordinary): 298.
\item[Q3)] Only frequently used lemmas: 9.
\item[Q4)] Frequently used lemmas and ordinary lemmas: 151.
\item[Q5)] Only ordinary lemmas: 230.
\end{enumerate}

Accordingly \cite{Veretennikov:IntelliSys:2018}, different algorithms will be applied to each kind of the query when multi-component key indexes are used.

For Q1 queries $(f, s, t)$ indexes are used and we have the following results.

Average query times:
$SE1$: 77.673 sec., $SE2.1$: 5.072 sec.,  $SE2.2$: 2.072 sec.,  $SE2.3$: 1.057 sec., and $SE2.4$: 0.662 sec.

Average data read sizes per query:
$SE1$: 2.027 GB, $SE2.1$: 190.6 MB, $SE2.2$: 57.4 MB, $SE2.3$: 33.5 MB, and $SE2.4$: 19.18 MB.

Average numbers of postings per query:
$SE1$: 511.5 million, $SE2.1$: 17.3 million, $SE2.2$: 5.08 million, $SE2.3$: 2.9 million, and $SE2.4$: 1.6 million.

We improved the average query processing time by a factor of 15.3 with the $SE2.1$ algorithm, by a factor of 37.4 with the $SE2.2$ algorithm, by a factor of 73.4 with the $SE2.3$ algorithm, and by a factor of 117.4 with the $SE2.4$ algorithm, in comparison with ordinary inverted files $SE1$.

We improved the average query processing time by a factor of 1.59 with the $SE2.4$ algorithm in comparison with the $SE2.3$ algorithm, by a factor of 3.1 in comparison with the $SE2.2$ algorithm, and by a factor of 7.6 in comparison with the $SE2.1$ algorithm.
 
For all Q1-Q5 queries we have the following results (with $SE2.4$ for Q1).

Average query times:
$Idx1$: 13.37 sec., $Idx2$: 0.521 sec.

Average data read sizes per query:
$Idx1$: 376 MB, $Idx2$: 12.85 MB.

Average numbers of postings per query:
$Idx1$: 90.2 million, $Idx2$: 0.81 million.

The average query processing time for multi-component key indexes is similar with Q1 case 
(e.g., with $Idx2$/SE2.4: 0.662 sec. for Q1 and 0.521 sec. for Q1-Q5),
but very different for ordinary inverted index $Idx1$ (77.673 sec. for Q1 and 13.37 sec. for Q1-Q5).
The multi-component key indexes allow to execute queries that consists of high-frequently occurring lemmas like ordinary queries in contradistinction with ordinary inverted indexes.

\paragraph{Example queries.}
Let us consider the following query: how to find the mean.

The query times:
$SE1$: 173.457 sec., $SE2.1$: 0.468 sec.,  $SE2.2$: 0.062 sec.,  $SE2.3$: 0.109 sec., and $SE2.4$: 0.094 sec.
The query can be executed with multi-component key indexes significantly faster than with the ordinary index $SE1$.

\paragraph{Duplicates.}
Although the average query times for $SE2.3$ and $SE2.4$ are near, 
it is important that the new algorithm $SE2.4$ can effective work with duplicates. 
Let us consider the following query: to be or not to be.
With $SE2.4$ this query was evaluated in 1.7 sec. and with $SE2.3$ in 10.1 sec. 
(the execution time above the average because lemmas of the query are very frequently occurring in texts).

\section{Incremental solution}
Let us consider the search query \textit{"Who I need you"} and the following text.

\noindent
\textit{The book that you are looking at is about the famous rock band ``The Who''. Their songs include ``I Need You'', ``You'', ``One at a Time'' and ``Who are you''.}

The partial tracing of the search is presented below. 
We use the following values in this example: $MaxDistance = 7$, $WindowSize = 14$.
The words are numbered, and these numbers are 1-based.

\noindent
Shift, $Start$ = 4 (start of Step 3, we use 4 to demonstrate the buffer switch).

\noindent
Read the posting (19, 20, 15), key (\textit{i}, \textit{need}, \textit{who}) (3.1).

\noindent
Set (position 19, key \textit{i}), buffer 1 (3.1).

\noindent
Set (position 20, key \textit{need}), buffer 1 (3.1).

\noindent
Set (position 15, key \textit{who}), buffer 0 (3.1).

\noindent
Read the posting (21, 20, 15), key (\textit{you}, \textit{need}*, \textit{who}*) (3.1).

\noindent
Set (position 21, key \textit{you}), buffer 1 (3.1).

\noindent
Read the posting (21, 20, 28), key (\textit{you}, \textit{need}*, \textit{who}*) (3.1).

\noindent
Set (position 21, key \textit{you}), buffer 1 (3.1).

\noindent
Read the posting (22, 20, 15), key (\textit{you}, \textit{need}*, \textit{who}*) (3.1).

\noindent
Set (position 22, key \textit{you}), buffer 1 (3.1).

\noindent
Read the posting (22, 20, 28), key (\textit{you}, \textit{need}*, \textit{who}*) (3.1).

\noindent
Set (position 22, key \textit{you}), buffer 1 (3.1).

\noindent
Populate the $Source$ queue using the data from the first buffer (3.1).

\noindent
Fetch (position 15, key \textit{who}) from the $Source$ queue (3.2, 3.3).

\noindent
Add (key \textit{who}) into $Lemma$ table, $Lemma.Count \neq Lemma.Max$ (3.4).

\noindent
Buffer switch, $Start$ = 18 (3.6).

\noindent
Populate the $Source$ queue using the data from the first buffer (3.1).

\noindent
Fetch (position 19, key \textit{i}) from the $Source$ queue (3.2, 3.3).

\noindent
Add (key \textit{i}) into $Lemma$ table, $Lemma.Count \neq Lemma.Max$ (3.4).

\noindent
Fetch (position 20, key \textit{need}) from the $Source$ queue (3.2, 3.3).

\noindent
Add (key \textit{need}) into $Lemma$ table, $Lemma.Count \neq Lemma.Max$ (3.4).

\noindent
Fetch (position 21, key \textit{you}) from the $Source$ queue (3.2, 3.3).

\noindent
Add (key \textit{you}) into $Lemma$ table, $Lemma.Count = Lemma.Max$ (3.4).

\noindent
Checking the $Lemma$ Table (3.5) $\rightarrow$ Result (from 15, to 21).

Please note that $WindowSize$ should be 64 for the better performance.

\section{Conclusion}

In the paper, we presented a new fast algorithm for proximity full-text search when a query that consists of high-frequently occurring words is considered. In the first experiment, we improved the average query processing time by a factor of 1.09 with the new algorithm in comparison with the algorithm from \cite{Veretennikov:DAMDID:2019}, by a factor of 1.3 in comparison with the algorithm from \cite{Veretennikov:DAMDID:2018} and by a factor of 1.5 in comparison with the algorithm from \cite{Veretennikov:SouthUral:2018}. This improvement can be done by using the requirement that we need to have a document which contains queried words near each other. We use three-component key indexes to solve the task.

In the second experiment (GOV2 text collection), we improved the average query processing time by a factor of 1.59 with the new algorithm in comparison with the algorithm from \cite{Veretennikov:DAMDID:2019}, by a factor of 3.1 in comparison with the algorithm from \cite{Veretennikov:DAMDID:2018}, and by a factor of 7.6 in comparison with the algorithm from \cite{Veretennikov:SouthUral:2018}.

We have presented the results of the experiments, showing that the average time of the query execution with our indexes is 142.13 times less (with a value of $MaxDistance = 5$) than that required when using ordinary inverted indexes, when queries that consist of high-frequently occurring words are evaluated. 

As we discussed in \cite{Veretennikov:SouthUral:2018}, three component key indexes occupy an important part of our holistic full-text search methodology. With them, we can evaluate queries that consist of high-frequently occurring words in an effective way. Other query types can be evaluated by using different additional indexes; those tasks are solved in \cite{Veretennikov:IntelliSys:2018} and, therefore, lie outside the scope of the current paper.

The new algorithm overcomes some limitations of our previous algorithm \cite{Veretennikov:DAMDID:2018,Veretennikov:DAMDID:2019}, that is, working with duplicate lemmas in the query and creating additional, considerable sized, intermediate data structures in the memory. 

In the future, it will be useful to optimize the index creation times for large values of $MaxDistance$. The new algorithm can also be used with any multi-component indexes and one-component indexes. The author is now creating indexes for relatively small values of $MaxDistance$, such as 5, 7, and 9.

The limitation of the proposed indexes is that we search only documents that contain queried words near each other. Documents in which the queried words occur at distances that are greater than $MaxDistance$ can be skipped. This limitation can be overcome by combining the proximity search with additional indexes with the search without distance \cite{Veretennikov:SouthUral:2018}. When the former requires a word-level index, the latter is requires only a document-level index. In the search without distance, we need only those documents that contain the queried words anywhere. This approach can produce fine results from the performance point of view if the documents are relatively large, e.g., several hundreds of kilobytes each. On the other hand, the modern approaches for calculating the relevance presuppose that the relevance of the document is inversely proportional to the square of the distance between searched words in the document \cite{Yan:2010:ETP:1871437.1871593}. Using this consideration, we can easily select a value of $MaxDistance$ that is large enough such that all relevant documents will be found by our additional indexes.

%
%


\begin{thebibliography}{6}
%

\bibitem{Anh:2001:VRE:383952.383957}
Anh, V.\,N., de~Kretser, O., Moffat, A.: Vector-space ranking with effective early termination. SIGIR '01 Proceedings of the 24th Annual International ACM SIGIR Conference on Research and Development in Information Retrieval, pp.~35--42. New Orleans, Louisiana, USA (2001). \url{doi:10.1145/383952.383957}

\bibitem{BorodinGIN}
Borodin, A., Mirvoda, S., Porshnev, S., Ponomareva, O.: Improving generalized inverted index lock wait times.
Journal of Physics: Conference Series, vol.~944, no.~1, Article number 012022 (2018). \url{doi:10.1088/1742-6596/944/1/012022}

\bibitem{Buttcher:2006:TPS:1148170.1148285}
B\"{u}ttcher, S., Clarke, C., Lushman, B.: Term proximity scoring for ad-hoc retrieval on very large text collections. SIGIR '06 Proceedings of the 29th annual international ACM SIGIR conference on Research and development in information retrieval, pp.~621--622 (2006). \url{doi:10.1145/1148170.1148285}

\bibitem{Daoud:2016:FTP:2978438.2978526}
Caio Moura Daoud, Silva de Moura, E., Carvalho, A., Soares da Silva, A., Fernandes, D., Rossi, C.: Fast top-k preserving query processing using two-tier indexes. Inf. Process. Manage, vol.~52, no.~5, pp.~855--872 (2016). \url{doi:10.1016/j.ipm.2016.03.005}

\bibitem{Fox:1989:SLG:378881.378888}
Fox, C.: A Stop List for General Text. ACM SIGIR Forum, vol.~24, pp.~19--35, (1989). \url{doi:10.1145/378881.378888}

\bibitem{Jansen:2000:RLR:342495.342498}
Jansen, B.\,J., Spink, A., Saracevic, T.: Real life, real users, and real needs: A study and analysis of user queries on the web. 
Inf. Process. Manage, vol.~36, no.~2, pp.~207--227 (2000). \url{doi:10.1016/S0306-4573(99)00056-4}

\bibitem{Jiang:2015:TEI:2839534.2840112}
Jiang, D., Kenneth Wai-Ting Leung, Yang, L., Ng, W.: TEII: topic enhanced inverted index for top-k document retrieval. Know.-Based Syst, vol.~89, no.~C., pp.~346--358 (2015). \url{doi:10.1016/j.knosys.2015.07.014}

\bibitem{GallKWordEncrypted}
Gall, M.,  Brost, G.: K-Word Proximity Search on Encrypted Data.
30th International Conference on Advanced Information Networking
and Applications Workshops (WAINA), pp. 365-372 (2016).
\url{doi:10.1 109/WAINA.2016.104}

\bibitem{Garcia:2004:AI:979922.979924}
Garcia, S., Williams, H.\,E., Cannane, A.: Access-ordered indexes. ACSC '04 Proceedings of the 27th Australasian Conference on Computer Science. Dunedin. New Zealand, pp.~7--14 (2004).

\bibitem{Lu:2016:EEH:2970398.2970404}
Lu, X., Moffat, A., Culpepper, J.S.: Efficient and effective higher order proximity modeling. ICTIR '16 Proceedings of the 2016 ACM International Conference on the Theory of Information Retrieval, pp.~21--30, (2016). \url{doi:10.1145/2970398.2970404}

\bibitem{Luk:2011:SSS:1988089.1988471}
Luk, R.\,W.\,P.: Scalable, statistical storage allocation for extensible inverted file construction. Journal of Systems and Software archive, vol.~84, no.~7, pp.~1082--1088 (2011). \url{doi:10.1016/j.jss.2011.01.049}

\bibitem{Sadakane1}
Sadakane, K.: Fast algorithms for k-word proximity search.
IEICE Transactions on Fundamentals of Electronics Communications and Computer Sciences,
vol. 84, no. 9, pp. 2311--2318 (2001).

\bibitem{Rasolofo:2003:TPS:1757788.1757808}
Rasolofo, Y., Savoy, J.: Term proximity scoring for keyword-based retrieval systems. European Conference on Information Retrieval (ECIR) 2003: Advances in Information Retrieval, pp.~207--218 (2003). \url{doi:10.1007/3-540-36618-0_15}

\bibitem{Veretennikov:DAMDID:2018}
Veretennikov, A.B.: Proximity full-text search with a response time guarantee by means of additional indexes with multi-component keys. Selected Papers of the XX International Conference on Data Analytics and Management in Data Intensive Domains (DAMDID/RCDL 2018), Moscow, Russia, October 9-12 2018, pp. 123--130 (2018). \url{http://ceur-ws.org/Vol-2277}


\bibitem{Veretennikov:DAMDID:2019}
Veretennikov, A.B.: Proximity Full-Text Search by Means of Additional Indexes with Multi-component Keys: In Pursuit of Optimal Performance.
In: Manolopoulos Y., Stupnikov S. (eds) Data Analytics and Management in Data Intensive Domains. DAMDID/RCDL 2018. Communications in Computer and Information Science, vol,~1003, pp.~111--130 (2019), Springer, Cham. \url{doi:10.1007/978-3-030-23584-0_7}

\bibitem{Veretennikov:IntelliSys:2018}
Veretennikov, A.B.: Proximity Full-Text Search with a Response Time Guarantee by Means of Additional Indexes. 
In: Arai K., Kapoor S., Bhatia R. (eds) Intelligent Systems and Applications. IntelliSys 2018. Advances in Intelligent Systems and Computing, vol.~868, pp.~936--954 (2019), Springer, Cham. \url{doi:10.1007/978-3-030-01054-6_66}

\bibitem{Veretennikov:SouthUral:2018}
Veretennikov, A.B.: Proximity full-text search with response time guarantee by means of three component keys.
Bulletin of the South Ural State University. Series: Computational Mathematics and Software Engineering, vol.~7, no.~1, pp.~60--77 (2018). (in Russian)

\bibitem{Williams:2004:FPQ:1028099.1028102}
Williams, H.\,E., Zobel, J., Bahle, D.: Fast phrase querying with combined indexes. ACM Transactions on Information Systems (TOIS).,  vol.~22, no.~4, pp.~573--594 (2004). \url{doi:10.1145/1028099.1028102}

\bibitem{Williams:HeapSort}
Williams, J.\,W.\,J.: Algorithm 232 – heapsort. Communications of the ACM, vol.~7, no.~6, pp.~347--348 (1964). \url{doi:10.2307/408772}

\bibitem{Yan:2010:ETP:1871437.1871593}
Yan, H., Shi, S., Zhang, F., Suel, T., Wen, J.-R.: Efficient term proximity search with term-pair indexes. CIKM '10 Proceedings of the 19th ACM International Conference on Information and Knowledge Management. Toronto, pp.~1229--1238 (2010). \url{doi:10.1145/1871437.1871593}

\bibitem{YangBlockLinked}
Yang, Y. Ning, H.:
Block linked list index structure for large data full text retrieval.
13th International Conference on Natural Computation, Fuzzy
Systems and Knowledge Discovery (ICNC-FSKD), pp 2123-2128. (2017).

\bibitem{Zipf:1929}
Zipf, G.: Relative frequency as a determinant of phonetic change. Harvard Studies in Classical Philology, vol.~40, pp.~1--95 (1929). \url{doi:10.2307/408772}

\bibitem{Zobel:2006:IFT:1132956.1132959}
Zobel, J., Moffat, A.: Inverted files for text search engines. ACM Comput. Surv., vol.~38, no.~2, Article 6 (2006). \url{doi:10.1145/1132956.1132959}


\end{thebibliography}
\end{document}